\documentclass[12pt]{article}
\usepackage[utf8]{inputenc}
\usepackage{natbib}

\usepackage{etoolbox}

\patchcmd{\thebibliography}{\bibsection}{\section*{References}}{}{}
\patchcmd{\thebibliography}{\section*{\refname}}{\section*{References}}{}{}
\patchcmd{\thebibliography}{\chapter*{\bibname}}{\section*{References}}{}{}

\usepackage{setspace}
\usepackage[english]{babel}
\usepackage{graphicx}
\usepackage{indentfirst}
\usepackage{framed}
\usepackage[normalem]{ulem}
\usepackage{amsmath}
\usepackage{amsthm}
\usepackage{amssymb}
\usepackage{amsfonts}
\usepackage{dcolumn}
\usepackage{enumerate}
\usepackage{float}
\usepackage{tabularx}
\usepackage{caption}
\usepackage{multirow}
\usepackage[utf8]{inputenc}
\usepackage{tabularx}
\usepackage{array}
\usepackage{float}
\usepackage{hyperref}
\usepackage{tikz}
\usepackage{pgf}
\usepackage[nottoc,numbib]{tocbibind}
\restylefloat{table}

\usepackage[top=1 in,bottom=1in, left=1 in, right=1 in]{geometry}

\newcolumntype{L}{>{\centering\arraybackslash}m{3cm}}

\theoremstyle{definition}

\newcolumntype{Y}{>{\centering\arraybackslash}X}

\begin{document}

\begin{center}
\Large{Adaptive CUSUM Chart for Simultaneous Monitoring of Mean and Variance}
\end{center}

\begin{center}
{\large Gokul Parakulum and Jun Li$^*$}\\
     Department of Statistics, University of California - Riverside\\
     $^*$Email: jun.li@ucr.edu.
\end{center}

\begin{abstract}
Simultaneously monitoring changes in both the mean and variance is a fundamental problem in Statistical Process Control, and numerous methods have been developed to address it. However, many existing approaches face notable limitations: some rely on tuning parameters that can significantly affect performance; others are biased toward detecting increases in variance while performing poorly for decreases; and some are computationally burdensome. To address these limitations, we propose a novel adaptive CUSUM chart for jointly monitoring the mean and variance of a Gaussian process. The proposed method is free of tuning parameters, efficient in detecting a broad range of shifts in both mean and variance, and well-suited for real-time monitoring due to its recursive structure. It also has a built-in post-signal diagnostics function that can identify
what kind of distributional changes have occurred after an alarm. Simulation results show that, compared to existing methods, the proposed chart achieves the most favorable balance between detection power and computational efficiency, delivering the best overall performance.
\end{abstract}

{\bf Key words:} Adaptive CUSUM chart; Simultaneous monitoring; Statistical process monitoring

\section{Introduction}

Statistical Process Control (SPC) emerged in the early 20th century as a pivotal methodology for maintaining product quality in manufacturing processes. Among its core tools are control charts, which were initially developed to detect shifts in a process’s mean. As SPC evolved, the importance of monitoring process variability became evident, prompting the development of charts specifically designed to detect changes in variance. However, in many real-world scenarios, shifts in both mean and variance can occur simultaneously or be closely interrelated. For example, an improperly positioned stencil in circuit  manufacturing can cause both a shift in the mean and an increase in the variance of the solder paste thickness applied to a circuit board (\cite{Gan2004}). Moreover, it is often unknown in advance whether a process disturbance will affect the mean, the variance, or both. This uncertainty underscores the need for joint monitoring schemes capable of detecting any meaningful changes in either parameter. 

A common approach is to monitor the mean and variance separately using traditional control charts such as the $\overline{X}$ chart for the mean and the $S$ or $R$ chart for variability (\cite{Gan1997}). While straightforward, these classical methods have several limitations. They often exhibit low sensitivity to small or moderate shifts and typically assume that the mean and variance can be monitored independently---an assumption that may not hold in practice.  Furthermore, separate monitoring can obscure important process changes. For instance, a simultaneous mean shift and variance reduction may go undetected because the decreased variability masks the mean shift, weakening the signal in the $\overline{X}$ chart (\cite{Hawkins2009}).

To overcome these drawbacks, numerous joint monitoring methods have been proposed. For example, \cite{Zhang2010} introduced a monitoring scheme based on the generalized likelihood ratio test (GLRT) for detecting simultaneous changes in the mean and variance. They combined this with exponentially weighted moving average (EWMA) estimators for the out-of-control (OC) parameters. Like other EWMA-based methods, their procedure depends critically on the choice of a weighting parameter $\lambda$: smaller values of $\lambda$ are more sensitive to small shifts, while larger values respond better to large shifts. However, since the nature of potential shifts is typically unknown in advance, poor tuning of $\lambda$ can substantially degrade performance, as confirmed by our simulation results.

To avoid the need of tuning parameters, \cite{Reynolds2013} proposed a GLRT-based chart that maximizes the likelihood over all possible change-points and OC parameter values. While this method is theoretically sound and tuning-free, it is computationally expensive, as it requires evaluating all possible change-point scenarios at each time step.

An alternative and increasingly influential approach is the adaptive CUSUM framework pioneered by \cite{Lorden2008}. This framework adaptively estimates the OC parameters while preserving the recursive nature of the CUSUM statistic, resulting in a computationally efficient and tuning-free method. It is also asymptotically optimal (up to second-order) for detecting changes in single-parameter exponential families.  \cite{Wu2017} extended this approach to multi-parameter exponential families and established its first-order asymptotic optimality. Since the Gaussian distribution is a special case of a two-parameter exponential family,  \cite{Wu2017} explicitly outlines a procedure for monitoring both the mean and variance in the Gaussian case. While his procedure can detect both increases and decreases in the mean, it can only detect increases in the variance. We argue that detecting decreases in variance is equally important for three key reasons. First, from a methodological perspective, a fully capable monitoring scheme should detect changes in any direction. Second, in manufacturing, early identification of quality improvement---reflected by reduced variability---can be also valuable.  Third, in emerging SPC applications such as healthcare, detecting variance decreases may be even more critical. For example, reduced heart rate variability (HRV) has been associated with dysfunctions in cardiovascular or autonomic nervous systems, making early detection of such decreases clinically significant (\cite{HRV_Johnston2019, HRVTurcu2023}).

To address the limitations of existing methods, we propose a new control chart for detecting real-time changes in the mean, variance, or both, of a Gaussian process. Our method also builds on the adaptive CUSUM framework due to its computational efficiency and optimality properties. Specifically, we extend the adaptive CUSUM procedure of \cite{Liu_etal2018}, originally designed to detect mean shifts alone, to also accommodate variance changes. Generalizing this approach introduces substantial challenges. In particular, robust detection of arbitrary changes requires maintaining eight distinct adaptive CUSUM statistics---each with its own in-control (IC) distribution. We address this complexity by introducing a principled aggregation strategy that fuses these statistics into a single, unified monitoring procedure. Simulation studies demonstrate that, compared to established alternatives in the literature, the proposed method offers a favorable balance: it delivers strong detection performance across a variety of shift types while remaining computationally efficient for real-time implementation.

The remainder of this paper is organized as follows.  
Section~2 describes the proposed adaptive CUSUM chart for joint monitoring of mean and variance in a Gaussian process.  
Section~3 presents simulation studies to evaluate the performance of the proposed procedure. Finally, Section~4 concludes with some closing remarks.

\section{Methodology}
\subsection{Problem Formulation and CUSUM Statistic}
In this paper, we consider the following setup. Suppose we observe a sequence of independent Gaussian random variables $X_1, X_2,...,X_t,...$. Initially, the process is in control, meaning that $X_t \sim N(\mu_0,\sigma_0^2)$, where $N(\mu_0,\sigma_0^2)$ denotes the IC normal distribution with mean $\mu_0$ and variance $\sigma_0^2$. After some unknown change-point $\tau$, the process shifts out of control, such that $X_t \sim N(\mu_1, \sigma_1^2)$ for $t>\tau$, where $\mu_1$ and $\sigma_1^2$ are the mean and variance of the OC distribution. 

At any given time $t$, we observe the sequence $X_1, X_2, \cdots, X_t$, and the task of a control chart is to determine, based on these observations, whether a change has occurred in the process. This corresponds to the following hypothesis testing problem:
\[
H_0: X_{1}, \cdots, X_t \text{ follow } N(\mu_0,\sigma_0^2),
\]
versus
\begin{equation}
\label{test0}
H_1: \exists \, \tau \in [1,t-1] \text{ s.t } X_1, \cdots, X_{\tau} \text{ follow } N(\mu_0,\sigma_0^2) \text{ and } X_{\tau+1}, \cdots, X_t \text{ follow } N(\mu_1, \sigma_1^2).
\end{equation}
As is standard in Phase II SPC, we assume that the IC parameters $\mu_0$ and $\sigma_0^2$ are either known or can be accurately estimated from a Phase I dataset. If we further assume that the OC parameters $\mu_1$ and $\sigma_1^2$  are also known, the log-likelihood ratio test statistic for the hypothesis testing problem in (\ref{test0}) is given by:
\[
C_{t}=\max(0,\max_{1 \leq \tau \leq t-1} \sum_{i=\tau+1}^t\log\left\{\frac{f_{1}(X_i)}{f_{0}(X_i)}\right\}),
\]
where $f_0$ and $f_1$ denote the probability density functions of $N(\mu_0,\sigma_0^2)$ and $N(\mu_1, \sigma_1^2)$, respectively. This statistic admits the following convenient recursive form:
\begin{equation}
\label{eqn:CUSUM}
C_{t}=\max(0,C_{t-1}+\log\left\{\frac{f_{1,X}(X_t)}{f_{0,X}(X_t)}\right\}),
\end{equation}
where $C_0=0$. This recursive formulation corresponds to the classic CUSUM statistic originally proposed by \cite{Page1954}. The resulting CUSUM chart is constructed by monitoring $C_{t}$ over time $t$ and signaling an alarm when $C_{t}$ exceeds a pre-specified threshold. Due to its simplicity and strong theoretical properties---including optimality results established by  \cite{Moustakides1986} ---the CUSUM chart has become one of the most widely used monitoring tools in SPC.

Substituting the explicit forms of the Gaussian densities  $f_{0}$ and $f_{1}$ into  (\ref{eqn:CUSUM}) yields:  
\[
C_{t} = \max\left(0, C_{t-1} + \frac{(X_{t}-\mu_{0})^{2}}{2\sigma_{0}^{2}} - \frac{(X_{t}-\mu_{1})^{2}}{2\sigma_{1}^{2}} -\frac{1}{2} \log\left(\frac{\sigma_{1}^{2}}{\sigma_{0}^{2}}\right) \right).
\]
Without loss of generality, since \( \mu_{0} \) and \( \sigma_{0} \) are known, we can standardize the process by setting \( \mu_{0} = 0 \) and \( \sigma_{0} = 1 \). This simplifies the expression to  
\begin{equation}
\label{eqn:CUSUM_normal}
C_{t} = \max\left(0, C_{t-1} +  \frac{X_{t}^{2}}{2} - \frac{(X_{t}-\mu_{1})^{2}}{2\sigma_{1}^{2}} - \frac{1}{2}\log(\sigma_{1}^{2})\right).
\end{equation}

This formulation represents the optimal CUSUM chart under the assumption that both OC parameters \( \mu_{1} \) and \( \sigma_{1}^{2} \) are known. However, in practice, these parameters are rarely available, and mis-specifying them can substantially degrade detection performance. This limitation motivates the development of adaptive monitoring procedures that do not require prior knowledge of the OC parameters, as explored in the remainder of this paper.

\subsection{Adaptive CUSUM Statistics}

From a theoretical point of view, there are two standard statistical approaches to handling unknown OC parameters. The first is the generalized likelihood ratio test (GLRT) approach, which maximizes the likelihood over all possible change-points and OC parameter values; see \cite{Lai:2001}. The GLRT-based chart proposed by \cite{Reynolds2013} follows this methodology. Although theoretically appealing, GLRT-based methods typically lack the recursive form enjoyed by CUSUM procedures, making them computationally intensive for real-time monitoring.

The second approach is the adaptive CUSUM framework, proposed by \cite{Lorden2008}. This framework retains the recursive structure of the CUSUM chart, offering computational efficiency while achieving asymptotic optimality under general conditions. Motivated by these advances, we adopt the adaptive CUSUM framework to jointly monitor changes in the mean and variance of a Gaussian process.

The key idea in adaptive CUSUM is to mimic the CUSUM statistic in~\eqref{eqn:CUSUM_normal}, but to replace the unknown OC parameters $\mu_1$ and $\sigma_1^2$ with their adaptive estimates. Before describing how to estimate these parameters, it is useful to note an important observation from the adaptive CUSUM procedure proposed by \cite{Liu_etal2018} for detecting mean shifts only: separate charts for increases and decreases in a parameter tend to be more effective than a single chart monitoring both directions simultaneously. Extending this principle to the joint monitoring of mean and variance, where each parameter may increase, decrease, or remain unchanged, leads to \(3^2 = 9\) possible combinations. Excluding the trivial (IC) case in which both parameters are unchanged, we obtain eight distinct OC scenarios.

To account for these possibilities, we maintain eight separate adaptive CUSUM statistics, one for each non-trivial combination of mean and variance changes. These are denoted as:
\[
C_{t}^{(+, +)}, \quad C_{t}^{(+, -)}, \quad C_{t}^{(-, +)}, \quad C_{t}^{(-, -)}, \quad C_{t}^{(., +)}, \quad C_{t}^{(., -)}, \quad C_{t}^{(+, .)}, \quad C_{t}^{(-, .)},
\]
where the superscript indicates the direction of change in the parameters \( (\mu, \sigma^2) \). Here, \( (+) \) denotes an increase, \( (-) \) a decrease, and \( (.) \) no change. For example, \( C_{t}^{(-, +)} \) corresponds to detecting a decrease in the mean and an increase in the variance.

An additional benefit of tracking these eight adaptive CUSUM statistics is interpretability: when an alarm is triggered, the specific adaptive CUSUM statistic that signals provides immediate insight into the nature of the shift, thus facilitating root-cause diagnosis. 

To compute each adaptive CUSUM statistic \( C_{t}^{(*_1, *_2)} \), where \( (*_1, *_2) \in \{+, -, .\}^2 \setminus \{(., .)\} \), we must estimate the OC parameters \( \mu_1 \) and \( \sigma_1^2 \) adaptively. If the change-point \( \tau \) were known, we could simply use post-change observations to estimate these parameters. However, \( \tau \) is unknown. To overcome this difficulty, we adopt the approach of \cite{Lorden2008} and \cite{Wu_ChangepointEstim_2005}, and define the estimated change-point at time \( t \), denoted by \( \widehat{\tau}_t^{(*_1,*_2)} \), as the most recent time at which the adaptive CUSUM statistic \( C_{t}^{(*_1, *_2)} \) resets to zero. Mathematically, \( \widehat{\tau}_t^{(*_1,*_2)} \) can be recursively updated as follows: $\widehat{\tau}_0^{(*_1,*_2)}=0$ and for $t \geq 1$, \[
\widehat{\tau}_t^{(*_1,*_2)}=
\begin{cases}
\widehat{\tau}_{t-1}^{(*_1,*_2)}, & \text{if } C_{t}^{(*_1, *_2)}>0,\\
t, & \text{if } C_{t}^{(*_1, *_2)}=0.
\end{cases}
\]
Based on this estimated change-point, we then can use observations in the interval \( [\widehat{\tau}_t^{(*_1,*_2)} + 1, t - 1] \) to estimate \( \mu_1 \) and \( \sigma_1^2 \), excluding the current observation \( X_t \) to avoid ``double dipping'' and to preserve both the recursive structure and martingale properties of \( C_{t}^{(*_1, *_2)} \) under the IC hypothesis. 

Define the number of observations in the interval \( [\widehat{\tau}_t^{(*_1,*_2)} + 1, t - 1] \) as:
\[
N_{t}^{(*_1,*_2)} = t - 1 - \widehat{\tau}^{(*_1,*_2)}.
\]
Based on how \( \widehat{\tau}_t^{(*_1,*_2)} \) is recursively updated, \(N_{t}^{(*_1,*_2)} \) can be also updated recursively by:
\begin{equation}
\label{eqn:Nt}
N_{t}^{(*_1,*_2)} =
\begin{cases}
N_{t-1}^{(*_1,*_2)} + 1, & \text{if } C_{t-1}^{(*_1, *_2)} > 0, \\    
0, & \text{if } C_{t-1}^{(*_1, *_2)} = 0,
\end{cases}   
\end{equation}
Based on $N_{t}^{(*_1,*_2)}$, we can see that, when \( t \) is close to \( \widehat{\tau}_t^{(*_1,*_2)} \), we have very few observations to estimate $\mu_1$ and $\sigma_1^2$. Although $\mu_1$ and $\sigma_1^2$  could, in principle, be estimated from as few as one or two observations, such estimates are highly variable. This variability can lead to excessively large values of \( C_{t}^{(*_1, *_2)} \), thus requiring larger control limits to maintain the desired IC average run length (ARL), which in turn reduces detection sensitivity (see, e.g., \cite{Reynolds2013}).

To mitigate this issue, we adopt a Bayesian approach, widely used in adaptive CUSUM procedures, which imposes a prior distribution on the OC parameters. This results in more stable and reliable estimates, especially when the number of available post-change observations is small.

\paragraph{Adaptive Estimation of the OC Mean}

To estimate \( \mu_1 \), we adopt the Bayesian estimators from \cite{Liu_etal2018}. For any \( *_1 \in \{+, -\} \) and \( *_2 \in \{+, -, .\} \), define
\[
S_{t}^{(*_1,*_2)} = \sum_{i = \widehat{\tau}^{(*_1,*_2)} + 1}^{t - 1} X_i,
\]
as the sum of observations in the interval \( [\widehat{\tau}_t^{(*_1,*_2)} + 1, t - 1] \). This quantity can be updated recursively by:
\begin{equation}
\label{eqn:St}
S_{t}^{(*_1,*_2)} =
\begin{cases}
S_{t-1}^{(*_1,*_2)} + X_{t-1}, & \text{if } C_{t-1}^{(*_1, *_2)} > 0, \\    
0, & \text{if } C_{t-1}^{(*_1, *_2)} = 0.
\end{cases}
\end{equation}

Given $S_{t}^{(*_1,*_2)}$ defined in (\ref{eqn:St}) and $N_{t}^{(*_1,*_2)}$ defined in (\ref{eqn:Nt}), the Bayesian estimators for \( \mu_1 \) corresponding to different types of changes in the mean are:
\begin{equation}
\label{eqn:mu1}
\widehat{\mu}_t^{(+, *_2)} = \max\left( \rho_{\mu}, \frac{s + S_t^{(+, *_2)}}{\eta + N_t^{(+, *_2)}} \right),
\end{equation}
\begin{equation}
\label{eqn:mu2}
\widehat{\mu}_t^{(-, *_2)} = \min\left( -\rho_{\mu}, \frac{-s + S_t^{(-, *_2)}}{\eta + N_t^{(-, *_2)}} \right),
\end{equation}
where \( \rho_{\mu} \) is a user-specified lower bound on the smallest meaningful shift in the mean (we set \( \rho_{\mu} = 0.25 \)), and \( s \), \( \eta \) are positive constants which act as prior parameters for the Bayesian estimator. For all settings in this paper, we set \( s = 1 \) and \( \eta = 4 \), which yields prior mean estimates of \( \pm 0.25 \) when \( C_{t}^{(*_1, *_2)} = 0 \), aligning with our interpretation of a minimal detectable shift.

\paragraph{Adaptive Estimation of the OC Variance}

To estimate the OC variance \( \sigma_1^2 \),  we employ a Bayesian approach by placing an inverse gamma prior with parameters $\alpha$ and $\beta$ (denoted by InverseGamma($\alpha,\beta$)) on the variance, as it serves as the conjugate prior for variance in the case of i.i.d. Gaussian data with known mean. For ease of notation, let \( \theta = \sigma^{2} \). For any \( *_1 \in \{+, -, .\} \) and \( *_2 \in \{+, -\} \), define
\[
Q_{t}^{(*_1,*_2)} = \sum_{i = \widehat{\tau}^{(*_1,*_2)} + 1}^{t - 1} \left(X_i-\widehat{\mu}_t^{(*_1,*_2)}\right)^2,
\]
as the sum of squared deviations of observations in the interval \( [\widehat{\tau}_t^{(*_1,*_2)} + 1, t - 1] \) from their adaptive mean estimate. In the above definition of $Q_{t}^{(*_1,*_2)}$, if  $*_1 \in \{+, -\}$, then $\widehat{\mu}_t^{(*_1,*_2)}$ is the adaptive estimate of $\mu_1$ as defined in (\ref{eqn:mu1}) or (\ref{eqn:mu2}); if $*_1=.$, then $\widehat{\mu}_t^{(*_1,*_2)}=0$, the IC mean, since $(.,*_2)$ denotes no change in the mean. The quantity $Q_{t}^{(*_1,*_2)}$ can be updated recursively by:
\[
Q_{t}^{(*_1,*_2)} =
\begin{cases}
   Q_{t-1}^{(*_1,*_2)} + (X_{t-1} - \widehat{\mu}^{(*_1,*_2)}_{t})^{2} & \text{if } C_{t-1}^{(*_1, *_2)} > 0\\    
   0 & \text{if } C_{t-1}^{(*_1, *_2)} = 0
\end{cases}
\]

Based on the above $Q_{t}^{(*_1,*_2)}$, the Bayesian estimator of $\theta$, corresponding to the posterior mean under the prior InverseGamma($\alpha,\beta$), is given by:
\[
\widehat{\theta}^{(*_1,*_2)} = \frac{\beta + Q_{t}^{(*_1,*_2)}/2}{\alpha -1+N_{t}^{(*_1,*_2)}/2},
\]
where $N_{t}^{(*_1,*_2)}$ is the effective sample size since the last estimated change-point as defined in (\ref{eqn:Nt}).

To ensure that the prior reflects the direction of variance changes each adaptive CUSUM statistic is designed to detect, we select the prior parameters $\alpha$ and $\beta$ so that the prior estimate of $\theta$ aligns with the expected direction of change. Specifically, we set the prior estimate of $\theta$ to be approximately \( 1.15 \) for variance increases and \( 1/1.15 \) for variance decreases. This is achieved by choosing the prior parameters as follows:
\[
\alpha^{(*_1,+)} = 12, \quad \beta^{(*_1,+)} = \beta^{(*_1,-)} = 15, \quad \alpha^{(*_1,-)} = 16.3.
\]
These parameter settings lead to the following Bayesian estimators for $\theta$, corresponding to upward and downward shifts in variance, respectively:  
\begin{equation}
\label{eqn:theta1}
\widehat{\theta}^{(*_1,+)} = \max\left(\rho^{+}_{\theta}, \frac{\beta^{(*_1,+)} + Q_{t}^{(*_1,+)}/2}{\alpha^{(*_1,+)} -1 + N_{t}^{(*_1,+)}/2} \right)
\end{equation}
\begin{equation}
\label{eqn:theta2}
\widehat{\theta}^{(*_1,-)} = \min\left(\rho^{-}_{\theta}, \frac{\beta^{(*_1,-)} + Q_{t}^{(*,-)}/2}{\alpha^{(*_1,-)} -1 + N_{t}^{(*_1,-)}/2} \right).
\end{equation}
As in the case with the mean estimates, to prevent variance increase estimates from falling below 1 and variance decrease estimates from rising above 1, we set $\rho_{\theta}^{+} = 1.05$ and $\rho_{\theta}^{-} = 1/1.05$ as the smallest meaningful variance shifts.

Finally, substituting the adaptive estimates of mean and variance from (\ref{eqn:mu1}), (\ref{eqn:mu2}), (\ref{eqn:theta1}) and (\ref{eqn:theta2}) into the Gaussian CUSUM statistic in~\eqref{eqn:CUSUM_normal}, we obtain the final adaptive CUSUM statistic:
\[
C_{t}^{(*_{1}, *_{2})} = \max\left(0, C^{(*_{1}, *_{2})}_{t-1} +  \frac{X_{t}^{2}}{2} - \frac{(X_{t}-\widehat{\mu}^{(*_{1}, *_{2})}_{t})^{2}}{2\widehat{\theta}_{t}^{(*_{1}, *_{2})}} - \frac{1}{2}\log(\widehat{\theta}_{t}^{(*_{1}, *_{2})}) \right),
\]
where \( C_{0}^{(*_{1}, *_{2})}=0\) and
\( (*_1, *_2) \in \{+, -, .\}^2 \setminus \{(., .)\} \).

\subsection{Aggregation of Adaptive CUSUM Statistics}

In the previous section, we described how to compute the adaptive CUSUM statistics at each time point. Since there are eight distinct adaptive CUSUM statistics, a key question arises: how should these eight adaptive CUSUM statistics be aggregated to assess the state of the process? Common aggregation strategies include summation or taking the maximum. However, before applying such methods, it is essential to ensure that the  statistics share a common IC distribution. Without this alignment, statistics with larger means or variances may dominate the combined measure, leading to biased or misleading decisions. Even if all statistics have the same mean and variance, differences in their underlying distributions can render them incomparable, as the same numerical value may correspond to different probabilities of the process being OC. In our case, the eight adaptive CUSUM statistics do not share a common IC distribution. This discrepancy was identified analytically at $t = 1$ and was confirmed via simulation for larger values of $t$. In particular, the adaptive CUSUM statistics $C_{t}^{(*_1,+)}$ for variance increases and  $C_{t}^{(*_1,-)}$ for variance decreases exhibit notably different IC distributions.

Due to their different IC distributions, direct comparison or aggregation of the raw adaptive CUSUM statistics is not viable. To circumvent this issue, we can find the IC distribution of each adaptive CUSUM statistic and then apply the probability integral transform to map them to a common distribution. Since an analytical derivation of the IC distribution is infeasible, we resort to empirical approximation at each time point. At first glance, this task appears daunting: the distribution of each statistic evolves over time and the time index \( t \) is unbounded. However, this challenge can be partially mitigated by noting that the IC evolution of each adaptive CUSUM statistic can be modeled as a Markov chain. In particular, the chain is irreducible, aperiodic, and positively recurrent, ensuring convergence to a unique stationary distribution. Therefore, it suffices to approximate the distribution during the early stages of evolution; beyond a certain point, the statistic can be treated as drawn from its stationary distribution.

An alternative approach is to initialize each adaptive CUSUM statistic and its associated estimators directly from their respective stationary (steady-state) distributions. This would ensure that all subsequent adaptive CUSUM statistics remain within the stationary regime, requiring storage only of their stationary IC distributions.

Define the IC cumulative distribution function (CDF) of the adaptive CUSUM statistic \( C_{t}^{(*_{1}, *_{2})} \), for \( (*_1, *_2) \in \{+,-,.\}^2 \setminus \{(., .)\} \), when the statistic is nonzero, as
\[
F_t^{(*_{1}, *_{2})}(c) := P^{(0)}\left(C_{t}^{(*_{1}, *_{2})} < c \,\middle|\, C_{t}^{(*_{1}, *_{2})} \neq 0\right).
\]
Suppose that we have a good empirical approximation of \( F_t^{(*_{1}, *_{2})} \), denoted by \( \widehat{F}_t^{(*_{1}, *_{2})} \), for all \( t \). Define the transformed statistic as
\[
p^{(*_{1}, *_{2})}_t = \widehat{F}_t^{(*_{1}, *_{2})}\left(C_{t}^{(*_{1}, *_{2})} \right).
\]
By the probability integral transform, it follows that \( p^{(*_{1}, *_{2})}_t \) follows a \( \text{Unif}(0,1) \) distribution under the IC condition, for all \( (*_1, *_2) \in \{+, -, .\}^2 \setminus \{(., .)\} \),  when the corresponding adaptive CUSUM statistic is nonzero. This transformation has a natural interpretation: when \( C_{t}^{(*_{1}, *_{2})} \) is close to 0, \( p^{(*_{1}, *_{2})}_{t} \) will also be close to 0. Conversely, when \( C_{t}^{(*_{1}, *_{2})} \) is large, \( p^{(*_{1}, *_{2})}_{t} \) will be close to 1. Using this interpretation, we aggregate the transformed statistics by taking the maximum:
\[
p_{t} = \max_{(*_1, *_2) \in \{+, -, .\}^2 \setminus \{(., .)\} } \left(p_{t}^{(*_1, *_2)}\right).
\]

Although one could monitor \( p_{t} \) directly to detect OC behavior, we instead apply a transformation to each \(p_{t}^{(*_1, *_2)}\) to obtain a positive, unbounded statistic, as this can simplify the estimation of control limits. Specifically, we define:
\[
q_{t}^{(*_1, *_2)}=-\log(1-p_{t}^{(*_1, *_2)}).
\]
By the inverse CDF transform, each \( q_{t}^{(*_1, *_2)} \) follows an exponential distribution with rate~1 under the IC condition when \( C_{t}^{(*_{1}, *_{2})} \neq 0 \). Accordingly, we aggregate these statistics as:
\[
q_{t} = \max_{(*_1, *_2) \in \{+, -, .\}^2 \setminus \{(., .)\} } \left(q_{t}^{(*_1, *_2)}\right).
\]
Our final adaptive CUSUM chart monitors the sequence \( q_t \) over time and raises an alarm if \( q_t \) exceeds a predefined threshold \( h \). This threshold \( h \) is selected to satisfy the desired IC ARL (denoted by \( ARL_0 \)) requirement and can be determined using a numerical method such as the bisection algorithm.

\subsection{Post-signal diagnostics}
When using a control chart to monitor both the mean and variance in practice, it is not only important to detect a change as quickly as possible, but also to identify the type of mean and/or variance change that has triggered the alarm. In the literature, most existing control charts for joint monitoring of the mean and variance require additional tests to determine the type of change after an alarm is raised. In contrast, our proposed adaptive CUSUM chart can automatically identify the change at the time the alarm is triggered.

To see this, recall that our adaptive CUSUM chart monitors
\[
q_{t} = \max_{(*_1, *_2) \in \{+, -, .\}^2 \setminus \{(., .)\} } \left(q_{t}^{(*_1, *_2)}\right),
\]
and raises an alarm whenever $q_t$ exceeds a control limit $h$. Because each $q_{t}^{(*_1, *_2)}$, with $(*_1, *_2) \in \{+, -, .\}^2 \setminus \{(., .)\}$, has the same IC run length distribution, our proposed monitoring scheme is equivalent to monitoring all $q_{t}^{(*_1, *_2)}$ separately and raising an alarm whenever at least one exceeds $h$. Each $q_{t}^{(*_1, *_2)}$ can be considered as a standardized version of the adaptive CUSUM statistic  $C_{t}^{(*_1, *_2)}$, where each $C_{t}^{(*_1, *_2)}$ is designed to detect a particular type of mean and/or variance change. Therefore, by checking which $q_{t}^{(*_1, *_2)}$ exceeds the control limit when the alarm is triggered, we can identify the specific type of change that caused the alarm. This provides a built-in post-signal diagnostic function, which is an additional advantage of our method.


\section{Simulation Studies}

In this section, we conduct simulation studies to evaluate the performance of our proposed adaptive CUSUM chart in detecting real-time changes in the mean, variance, or both, of a Gaussian process. Specifically, we compare our method against several existing monitoring procedures designed for the same purpose. Below, we briefly review the benchmark methods used for comparison.

\subsection{Review of Benchmark Methods}
\noindent \textbf{EWMA-based GLRT Chart by \cite{Zhang2010}}

\cite{Zhang2010} proposed a monitoring statistic based on the GLRT for detecting simultaneous changes in the mean and variance of a Gaussian distribution. In general, the GLRT statistic for a random sample $x_1,...,x_n$ is given by:  
\[
\Lambda = \sup_{\mu,\sigma^2 \in \mathbb{R}\times \mathbb{R}^+}\prod_{i=1}^{t}\frac{f(x_i; \mu, \sigma^2)}{f(x_i; 0,1)}.
\]
For Gaussian data, this reduces to 
\begin{equation*}
\log(\Lambda) = \frac{n}{2} \left[ \bar{x}^2 + \hat{\sigma}_1^2 - \log(\hat{\sigma}_1^2) - 1 \right],
\end{equation*}
where \( \bar{x} = \frac{1}{n} \sum_{i=1}^n x_i \) is the sample mean, and \( \hat{\sigma}^2 = \frac{1}{n} \sum_{i=1}^n (x_i - \bar{x})^2 \) is the maximum likelihood estimator (MLE) of the variance.

To adapt this statistic to the online setting, \cite{Zhang2010} replaced \( \bar{x} \) and \( \hat{\sigma}^2 \) with EWMA estimators. Specifically, they defined:
\[
u_t = \lambda x_t + (1 - \lambda) u_{t-1}, \quad
v_t = \lambda (x_t - u_t)^2 + (1 - \lambda) v_{t-1},
\]
where \( u_0 = 0 \) and \( v_0 = 1 \), and \( \lambda \in (0,1) \) is a weighting parameter. The resulting online monitoring statistic is:
\[
ELR_t = u_t^2 + v_t - \log(v_t) - 1,
\]
which mimics the structure of the fixed-sample GLRT. An alarm is raised when \( ELR_t \) exceeds a pre-specified control limit.
\vspace{4mm}

\noindent \textbf{GLRT Chart by \cite{Reynolds2013}}

At any given time \( t \), the GLRT chart is based on the following likelihood ratio:
\[
\Lambda_t = \sup_{\substack{\tau \in \{1, \ldots, t-1\} \\ \mu, \sigma^2 \in \mathbb{R} \times \mathbb{R}^+}} 
\prod_{i=\tau + 1}^{t} \frac{ f(x_i ; \mu, \sigma^2) }{ f(x_i; 0, 1) }.
\]
The GLRT chart is constructed by monitoring \( \log(\Lambda_t) \), and an alarm is raised when \( \log(\Lambda_t) \) exceeds a pre-specified control limit.

\cite{Reynolds2013} applied this methodology to monitor changes in both mean and variance of a Gaussian process. However, a known limitation of the GLRT chart is its instability when the MLEs are computed from a small number of observations---specifically, when \( \tau \) is close to \( t - 1 \). This issue is particularly acute for variance estimation, where the MLE can become overly sensitive to small sample sizes.

To address this, \cite{Reynolds2013} proposed modifying the variance estimator in the likelihood ratio. Rather than using the MLE, they introduced a stabilized estimator defined as:
\begin{equation}
    \sigma^2_{B,\tau} := \max \left( 1 - \gamma(t - \tau), \, \hat{\sigma}^2_{\tau} \right),
    \label{eq: GLRT variance estimate}
\end{equation}
where \( \hat{\sigma}^2_{\tau} \) is the usual sample variance based on observations \( \{x_i\}_{i = \tau+1}^{t} \), and \( \gamma \) is a tuning parameter. This adjustment imposes a lower bound on the variance estimate, which becomes more stringent when fewer observations are available, thereby preventing unrealistically small estimates that could distort the likelihood ratio. The MLE for the mean remains unchanged in their formulation.

\vspace{4mm}

\noindent \textbf{Adaptive CUSUM Chart by \cite{Wu2017}}

\cite{Wu2017} outlines a procedure for monitoring changes in both mean and variance in the Gaussian setting. While the procedure can detect both increases and decreases in the mean, it is designed to detect only increases in the variance. \cite{Wu2017} also recommends using separate charts to monitor increases and decreases in the mean. Accordingly, we denote his chart for detecting mean increases as \( T_t^+ \) and that for mean decreases as \( T_t^- \). Let \( * \in \{+, -\} \), and let \( \theta \) denote the variance \( \sigma^2 \) for notational convenience. The adaptive CUSUM procedure in \cite{Wu2017} consists of the following steps:

\begin{itemize}
    \item \textbf{Step 1:} Initialize the procedure with
    \[
    T_0^* = 0, \quad \tau_0^* = 0, \quad \hat{\mu}_0^* = \delta^*, \quad \hat{\theta}_0^* = \rho.
    \]

    \item \textbf{Step 2:} Update the monitoring statistic as
    \[
    T_t^* = \max\left\{ 0, T^*_{t-1} + \frac{1}{2}\left( \frac{\hat{\theta}^*_{t-1} - 1}{\hat{\theta}^*_{t-1}} X_t^2 - \log(\hat{\theta}^*_{t-1}) \right) + \frac{\hat{\mu}^*_{t-1}}{\hat{\theta}^*_{t-1}} \left( X_t - \frac{\hat{\mu}^*_{t-1}}{2} \right) \right\}.
    \]

    \item \textbf{Step 3:} If \( T_t^* = 0 \), reset the parameters:
    \[
    \tau_t^* = t, \quad \hat{\mu}_t^* = \delta^*, \quad \hat{\theta}_t^* = \rho.
    \]
    Otherwise, update the estimates as:
    \[
    \hat{\mu}_t^* = \max\left\{ \delta^*, \hat{\mu}_{t-1}^* + \frac{1}{a + t - \tau_t^*}(X_t - \hat{\mu}_{t-1}^*) \right\},
    \]
    \[
    \hat{\theta}_t^* = \max\left\{ \rho, \hat{\theta}_{t-1}^* + \frac{1}{b + t - \tau_t^*}\left[(X_t - \hat{\mu}_{t-1}^*)^2 - \hat{\theta}_{t-1}^* \right] \right\}.
    \]

    \item \textbf{Step 4:} Raise an alarm when \( \max(T_t^+, T_t^-) \) exceeds a pre-specified control limit.
\end{itemize}
In the above procedure, the constants \( a \) and \( b \) serve as tuning parameters, while \( \delta^* \) and \( \rho\) represent the user-specified minimum detectable shifts in the mean and variance, respectively. In \cite{Wu2017}, the following values were used:  \( a=b=0.5 \),  \( \delta^+ = 0.5 \), \( \delta^- = -0.5 \) and \( \rho=1.05\). In our proposed procedure, we set the minimum detectable mean shift to \( \pm 0.25 \). For a fair comparison in our simulation study, we also set \( \delta^* = \pm 0.25 \).

\subsection{Simulation Settings}

We conduct a series of simulation studies to evaluate and compare the detection performance of our proposed adaptive CUSUM chart (denoted by \texttt{P\_CUSUM}) against the three benchmark procedures described in the previous subsection:
\begin{itemize}
    \item The adaptive CUSUM chart proposed by \cite{Wu2017}, denoted by \texttt{Wu\_CUSUM}.
    \item The GLRT chart proposed by \cite{Reynolds2013}, denoted by \texttt{GLRT}.
    \item The EWMA-based GLRT chart by \cite{Zhang2010}, denoted by \texttt{EWMA\_GLRT}.
\end{itemize}

Results are presented in six separate tables, each corresponding to a specific type of distributional change, as outlined below:
\begin{enumerate}
    \item \textbf{Mean shifts only}: Only the mean changes.
    \item \textbf{Variance shifts only}: Only the variance changes.
    \item \textbf{Simultaneous shifts with variance increases}: Both the mean and variance change, with variance increasing. 
    \item \textbf{Simultaneous shifts with variance decreases}: Both the mean and variance change, with variance decreasing.
\end{enumerate}

These scenarios comprehensively cover the space of potential shifts in the mean and/or variance, while maintaining clarity and organization. In all tables, the first row corresponds to the IC setting, which is assumed---without loss of generality---to be \((\mu, \sigma) = (0, 1)\). The first column, labeled \((\mu, \sigma)\), specifies the OC distribution in each case.

For the \texttt{GLRT} chart, two tuning parameters are required: the window size and the lower-bound parameter \(\gamma\) used in the variance estimator (see (\ref{eq: GLRT variance estimate})). Following the authors' recommendations, we set the window size to 800. For \(\gamma\), the original paper suggests values of 0.001, 0.0025, and 0.005. Since our evaluation assigns equal importance to detecting variance decreases, we choose \(\gamma = 0.005\), which yielded the best performance in that direction.

The performance of the \texttt{EWMA\_GLRT} chart is reported in the final four columns of each table, corresponding to four widely used values of the weighting parameter \(\lambda\). Given the sensitivity of EWMA performance to \(\lambda\), the best-performing value for each scenario is highlighted in bold. Note that the \texttt{Wu\_CUSUM} chart is not designed to detect variance decreases and is therefore omitted from the relevant comparison in Table 2.

To simplify implementation, rather than empirically estimating the time-varying IC distributions of our proposed adaptive CUSUM statistics until they stabilize, we initialize each statistic and its associated estimators using samples drawn directly from their respective stationary (steady-state) distributions. This approach ensures that all statistics operate within the stationary regime from the outset, eliminating the need to store the full time-varying distributions. Furthermore, starting from steady state naturally yields steady-state performance results, which are commonly used in control chart evaluations.

All simulations were conducted with \(ARL_0=500\) and the change-point fixed at \(\tau = 50\). For each scenario, 10{,}000 simulation runs were performed. Following standard practice, if a chart signaled before the change-point, the run was discarded and re-simulated until a valid post-change signal was observed. The tables report the post-change ARL (denoted by \(ARL_1\)), with standard errors shown in parentheses.

\subsection{Simulation Results}

\begin{table}[!htpb]
\resizebox{\textwidth}{!}{
\def\arraystretch{1.5}
\centering
\begin{tabular}{llllllll}
\hline
\hline
 & & & & \multicolumn{4}{c}{EWMA\_GLRT} \\
 \cline{5-8}
\((\mu, \sigma)\) & P\_CUSUM & Wu\_CUSUM & GLRT & $\lambda = 0.01$ & $\lambda = 0.05$ & $\lambda = 0.1$ & $\lambda = 0.2$ \\
\hline
\hline
(0, 1) & 496.4 (5.11) & 500.1 (4.94) & 494.8 (4.67) & 501.4 (4.52) & 508.7 (4.84) & 502.6 (4.85) & 500.7 (4.86) \\ 
(0.25, 1) & 101.3 (0.83) & 130.6 (1.18) & 101.3 (0.755) & \textbf{85 (0.521)} & 99 (0.832) & 124.9 (1.15) & 162.4 (1.58) \\ 
(0.5, 1) & 31.9 (0.212) & 35.9 (0.272) & 33.9 (0.214) & 38.2 (0.188) & \textbf{32.9 (0.204)} & 36.4 (0.273) & 45.4 (0.403) \\ 
(0.75, 1) & 16.5 (0.093) & 17 (0.111) & 17.5 (0.103) & 23.9 (0.105) & 18.2 (0.095) & \textbf{17.7 (0.103)} & 19.8 (0.145) \\ 
(1, 1) & 10.7 (0.053) & 10.3 (0.06) & 10.9 (0.06) & 16.6 (0.072) & 12.3 (0.057) & \textbf{11.4 (0.058)} & 11.6 (0.069) \\ 
(1.25, 1) & 7.6 (0.035) & 7.2 (0.039) & 7.5 (0.039) & 12.3 (0.054) & 9.1 (0.041) & 8.2 (0.038) & \textbf{7.9 (0.042)} \\ 
(1.5, 1) & 6 (0.026) & 5.4 (0.026) & 5.6 (0.028) & 9.5 (0.04) & 7 (0.031) & 6.4 (0.029) & \textbf{5.9 (0.029)} \\ 
(1.75, 1) & 4.9 (0.019) & 4.4 (0.02) & 4.4 (0.021) & 7.6 (0.032) & 5.6 (0.025) & 5.1 (0.022) & \textbf{4.8 (0.022)} \\ 
(2, 1) & 4.1 (0.015) & 3.7 (0.015) & 3.6 (0.016) & 6.3 (0.026) & 4.7 (0.02) & 4.2 (0.018) & \textbf{3.8 (0.017)} \\ 
\hline
\end{tabular}}
\caption{Performance comparison for mean shifts only.}
\end{table}

\begin{table}[!htpb]
\resizebox{\textwidth}{!}{
\def\arraystretch{1.5}
\centering
\begin{tabular}{llllllll}
\hline
\hline
 & & & & \multicolumn{4}{c}{EWMA\_GLRT} \\
 \cline{5-8}
\((\mu, \sigma)\) & P\_CUSUM & Wu\_CUSUM & GLRT & $\lambda = 0.01$ & $\lambda = 0.05$ & $\lambda = 0.1$ & $\lambda = 0.2$ \\
\hline
\hline
(0, 1) & 495.6 (5.09) & 499.6 (4.9) & 488.1 (4.64) & 498.5 (4.46) & 512.8 (4.88) & 510.5 (5.01) & 486.8 (4.78) \\ 
  (0, 1.1) & 198.6 (1.85) & 229.6 (2.23) & 179.6 (1.65) & \textbf{166.5 (1.39)} & 228.7 (2.26) & 265.9 (2.63) & 281.7 (2.82) \\ 
  (0, 1.15) & 122.3 (1.08) & 160.9 (1.55) & 115.2 (1.02) & \textbf{103.2 (0.775)} & 136.1 (1.28) & 172.4 (1.69) & 197.1 (1.94) \\ 
  (0, 1.2) & 83.4 (0.687) & 121.3 (1.18) & 78.1 (0.667) & \textbf{75.5 (0.534)} & 90.3 (0.819) & 114.9 (1.1) & 136.7 (1.35) \\ 
  (0, 1.3) & 47.2 (0.36) & 73.3 (0.699) & 43.7 (0.36) & \textbf{45.6 (0.287)} & 46.4 (0.387) & 57 (0.527) & 73.1 (0.712) \\ 
  (0, 1.4) & 31.4 (0.226) & 50.1 (0.465) & 28.6 (0.227) & 32.1 (0.194) & \textbf{30.4 (0.232)} & 34.9 (0.305) & 44.4 (0.421) \\ 
  (0, 1.6) & 17.8 (0.123) & 28 (0.25) & 15.8 (0.123) & 19.9 (0.118) & \textbf{17.1 (0.124)} & 17.7 (0.144) & 21.2 (0.191) \\ 
  (0, 1.8) & 12.4 (0.082) & 18.8 (0.168) & 10.6 (0.083) & 14.1 (0.085) & 11.6 (0.08) & \textbf{11.5 (0.089)} & 13 (0.113) \\ 
  (0, 2) & 9.3 (0.061) & 14 (0.122) & 7.7 (0.06) & 10.9 (0.066) & 8.8 (0.06) & \textbf{8.6 (0.064)} & 9.1 (0.075) \\ 
  (0, 0.9) & 249.5 (1.9) & - (-) & 283.1 (2.05) & \textbf{162.7 (1.13)} & 227.1 (2.09) & 303.1 (2.94) & 427.5 (4.22) \\ 
  (0, 0.85) & 140.6 (0.881) & - (-) & 160.1 (0.95) & \textbf{104.6 (0.587)} & 128.5 (1.06) & 183.4 (1.7) & 311 (3.02) \\ 
  (0, 0.8) & 92.1 (0.472) & - (-) & 103.6 (0.509) & \textbf{76.7 (0.361)} & 80.3 (0.594) & 112.8 (0.975) & 203.1 (1.95) \\ 
  (0, 0.7) & 53.4 (0.196) & - (-) & 61.3 (0.185) & 51.2 (0.195) & \textbf{41.1 (0.213)} & 49.1 (0.336) & 85.8 (0.764) \\ 
  (0, 0.6) & 38.4 (0.113) & - (-) & 47.8 (0.112) & 39.4 (0.135) & \textbf{27.4 (0.107)} & 27.4 (0.136) & 39.9 (0.298) \\ 
  (0, 0.5) & 30.5 (0.076) & - (-) & 41.3 (0.091) & 32.8 (0.107) & 21.2 (0.071) & \textbf{19 (0.07)} & 21.5 (0.119) \\ 
  (0, 0.4) & 26.3 (0.058) & - (-) & 37.1 (0.082) & 29 (0.092) & 17.7 (0.055) & 14.9 (0.045) & \textbf{14.1 (0.053)} \\ 
  (0, 0.3) & 23.7 (0.05) & - (-) & 34.6 (0.076) & 26.5 (0.085) & 15.8 (0.049) & 12.7 (0.036) & \textbf{10.8 (0.031)} \\ 
  \hline
\end{tabular}}
\caption{Performance comparison for variance shifts only.}
\end{table}

\begin{table}[!htpb]
\resizebox{\textwidth}{!}{
\def\arraystretch{1.5}
\centering
\begin{tabular}{llllllll}
\hline
\hline
 & & & & \multicolumn{4}{c}{EWMA\_GLRT} \\
 \cline{5-8}
\((\mu, \sigma)\) & P\_CUSUM & Wu\_CUSUM & GLRT & $\lambda = 0.01$ & $\lambda = 0.05$ & $\lambda = 0.1$ & $\lambda = 0.2$ \\
\hline
\hline
(0, 1) & 495.9 (5.07) & 502.5 (4.88) & 504.3 (4.78) & 500 (4.42) & 504.2 (4.93) & 504.1 (4.86) & 506.3 (4.94) \\  
  (0.25, 1.1) & 79.9 (0.673) & 91.8 (0.853) & 76.8 (0.613) & \textbf{70.7 (0.462)} & 79.6 (0.685) & 96.5 (0.897) & 121.2 (1.19) \\  
  (0.5, 1.1) & 30 (0.212) & 32.5 (0.26) & 30.5 (0.212) & 35.2 (0.185) & \textbf{31 (0.207)} & 33.4 (0.257) & 40.5 (0.357) \\  
  (0.75, 1.1) & 16.1 (0.097) & 16.4 (0.115) & 16.2 (0.104) & 22.1 (0.107) & 17.4 (0.098) & \textbf{17.2 (0.11)} & 18.9 (0.144) \\  
  (1, 1.1) & 10.5 (0.057) & 10.4 (0.065) & 10.4 (0.063) & 15.7 (0.071) & 11.8 (0.06) & \textbf{11.2 (0.062)} & 11.3 (0.073) \\  
  (1.25, 1.1) & 7.7 (0.038) & 7.1 (0.04) & 7.3 (0.042) & 11.7 (0.054) & 8.8 (0.043) & 8.1 (0.042) & \textbf{7.9 (0.046)} \\  
  (1.5, 1.1) & 6 (0.027) & 5.5 (0.028) & 5.5 (0.029) & 9.2 (0.042) & 6.9 (0.032) & 6.3 (0.031) & \textbf{5.9 (0.031)} \\  
  (1.75, 1.1) & 4.8 (0.021) & 4.5 (0.021) & 4.4 (0.022) & 7.4 (0.033) & 5.5 (0.025) & 5 (0.023) & \textbf{4.7 (0.023)} \\  
  (2, 1.1) & 4.1 (0.017) & 3.7 (0.016) & 3.6 (0.017) & 6.1 (0.027) & 4.6 (0.021) & 4.2 (0.019) & \textbf{3.8 (0.018)} \\  
  (0.25, 1.2) & 53.8 (0.434) & 67.5 (0.624) & 50.8 (0.411) & \textbf{51.4 (0.328)} & 53.5 (0.442) & 64.2 (0.583) & 78.5 (0.754) \\  
  (0.5, 1.2) & 26.5 (0.196) & 29.4 (0.243) & 26 (0.191) & 30.8 (0.168) & \textbf{26.9 (0.183)} & 28.7 (0.226) & 33.3 (0.291) \\  
  (0.75, 1.2) & 15.3 (0.099) & 15.9 (0.116) & 15.1 (0.103) & 20.4 (0.104) & \textbf{16.3 (0.098)} & \textbf{16.3 (0.111)} & 17.2 (0.133) \\  
  (1, 1.2) & 10.3 (0.059) & 10.2 (0.068) & 10 (0.063) & 14.8 (0.073) & 11.3 (0.061) & \textbf{10.6 (0.063)} & 10.8 (0.074) \\  
  (1.25, 1.2) & 7.5 (0.039) & 7.3 (0.044) & 7.2 (0.043) & 11.2 (0.053) & 8.5 (0.044) & 7.9 (0.043) & \textbf{7.6 (0.045)} \\  
  (1.5, 1.2) & 5.9 (0.029) & 5.6 (0.031) & 5.5 (0.031) & 8.9 (0.042) & 6.7 (0.034) & 6.1 (0.031) & \textbf{5.8 (0.032)} \\  
  (1.75, 1.2) & 4.8 (0.022) & 4.5 (0.022) & 4.3 (0.024) & 7.2 (0.033) & 5.4 (0.026) & 4.9 (0.025) & \textbf{4.6 (0.024)} \\  
  (2, 1.2) & 4.1 (0.018) & 3.7 (0.017) & 3.5 (0.019) & 5.9 (0.027) & 4.5 (0.021) & 4.1 (0.02) & \textbf{3.9 (0.019)} \\  
  (0.25, 1.4) & 26.9 (0.196) & 38.7 (0.353) & 24.4 (0.194) & 28.3 (0.171) & \textbf{26.2 (0.197)} & 28.5 (0.242) & 34.9 (0.322) \\  
  (0.5, 1.4) & 18.8 (0.133) & 22.9 (0.194) & 17.2 (0.131) & 21.9 (0.125) & \textbf{18.8 (0.131)} & 19.4 (0.149) & 21.9 (0.188) \\  
  (0.75, 1.4) & 13 (0.087) & 14.7 (0.117) & 12 (0.088) & 16.6 (0.091) & 13.4 (0.085) & \textbf{13.2 (0.094)} & 14 (0.114) \\  
  (1, 1.4) & 9.5 (0.059) & 9.9 (0.072) & 8.7 (0.062) & 12.8 (0.067) & 10 (0.06) & \textbf{9.3 (0.061)} & 9.5 (0.071) \\  
  (1.25, 1.4) & 7.2 (0.041) & 7.3 (0.048) & 6.6 (0.043) & 10 (0.052) & 7.8 (0.045) & 7.2 (0.043) & \textbf{7 (0.047)} \\  
  (1.5, 1.4) & 5.8 (0.031) & 5.7 (0.034) & 5.2 (0.032) & 8.2 (0.042) & 6.2 (0.034) & 5.8 (0.033) & \textbf{5.5 (0.034)} \\  
  (1.75, 1.4) & 4.8 (0.024) & 4.6 (0.026) & 4.2 (0.025) & 6.8 (0.034) & 5.2 (0.028) & 4.7 (0.027) & \textbf{4.5 (0.026)} \\  
  (2, 1.4) & 4.1 (0.019) & 3.9 (0.02) & 3.5 (0.02) & 5.7 (0.029) & 4.4 (0.023) & 4 (0.021) & \textbf{3.8 (0.021)} \\  
    \hline
\end{tabular}}
\caption{Performance comparison for simultaneous shifts with variance increases: Part 1.}
\end{table}

\begin{table}[!htpb]
\resizebox{\textwidth}{!}{
\def\arraystretch{1.5}
\centering
\begin{tabular}{llllllll}
\hline
\hline
 & & & & \multicolumn{4}{c}{EWMA\_GLRT} \\
 \cline{5-8}
\((\mu, \sigma)\) & P\_CUSUM & Wu\_CUSUM & GLRT & $\lambda = 0.01$ & $\lambda = 0.05$ & $\lambda = 0.1$ & $\lambda = 0.2$ \\
\hline
\hline
(0, 1) & 495.9 (5.07) & 502.5 (4.88) & 504.3 (4.78) & 500 (4.42) & 504.2 (4.93) & 504.1 (4.86) & 506.3 (4.94) \\  
  (0.25, 1.7) & 13.9 (0.095) & 20.5 (0.179) & 11.8 (0.093) & 15.8 (0.092) & \textbf{13.3 (0.093)} & \textbf{13.3 (0.104)} & 15.1 (0.131) \\  
  (0.5, 1.7) & 12 (0.081) & 16 (0.137) & 10.3 (0.079) & 14.2 (0.082) & 11.4 (0.077) & \textbf{11.3 (0.084)} & 12.3 (0.103) \\  
  (0.75, 1.7) & 9.9 (0.066) & 11.8 (0.097) & 8.6 (0.064) & 11.9 (0.069) & 9.6 (0.064) & \textbf{9.3 (0.067)} & 9.8 (0.079) \\  
  (1, 1.7) & 7.9 (0.051) & 8.9 (0.068) & 6.9 (0.051) & 10.1 (0.058) & 7.9 (0.05) & \textbf{7.5 (0.051)} & 7.6 (0.058) \\  
  (1.25, 1.7) & 6.5 (0.04) & 7 (0.051) & 5.6 (0.04) & 8.5 (0.048) & 6.6 (0.04) & 6.2 (0.041) & \textbf{6 (0.043)} \\  
  (1.5, 1.7) & 5.5 (0.032) & 5.7 (0.038) & 4.6 (0.032) & 7.2 (0.04) & 5.5 (0.034) & 5.1 (0.033) & \textbf{5 (0.034)} \\  
  (1.75, 1.7) & 4.6 (0.026) & 4.7 (0.03) & 3.9 (0.025) & 6.1 (0.034) & 4.7 (0.028) & 4.4 (0.026) & \textbf{4.2 (0.027)} \\  
  (2, 1.7) & 4.1 (0.022) & 4 (0.024) & 3.4 (0.021) & 5.2 (0.028) & 4.1 (0.024) & 3.8 (0.022) & \textbf{3.6 (0.023)} \\  
  (0.25, 2) & 9.2 (0.062) & 13.4 (0.117) & 7.5 (0.058) & 10.5 (0.063) & 8.6 (0.058) & \textbf{8.2 (0.061)} & 8.8 (0.072) \\  
  (0.5, 2) & 8.5 (0.056) & 11.6 (0.099) & 7 (0.052) & 9.9 (0.059) & 8 (0.053) & \textbf{7.7 (0.057)} & 8 (0.065) \\  
  (0.75, 2) & 7.5 (0.049) & 9.7 (0.08) & 6.2 (0.047) & 8.9 (0.053) & 7.2 (0.048) & \textbf{6.8 (0.049)} & 7 (0.056) \\  
  (1, 2) & 6.6 (0.042) & 8 (0.064) & 5.5 (0.04) & 7.9 (0.047) & 6.4 (0.042) & \textbf{6 (0.041)} & 6.1 (0.047) \\  
  (1.25, 2) & 5.8 (0.036) & 6.5 (0.049) & 4.9 (0.035) & 7 (0.042) & 5.6 (0.036) & \textbf{5.2 (0.036)} & \textbf{5.2 (0.038)} \\  
  (1.5, 2) & 5.1 (0.031) & 5.6 (0.039) & 4.2 (0.029) & 6.3 (0.037) & 4.9 (0.031) & \textbf{4.5 (0.03)} & \textbf{4.5 (0.032)} \\  
  (1.75, 2) & 4.4 (0.026) & 4.8 (0.033) & 3.7 (0.025) & 5.5 (0.032) & 4.3 (0.027) & 4 (0.027) & \textbf{3.9 (0.027)} \\  
  (2, 2) & 3.9 (0.022) & 4.2 (0.027) & 3.2 (0.021) & 4.8 (0.028) & 3.8 (0.023) & 3.5 (0.022) & \textbf{3.4 (0.023)} \\  
    \hline
\end{tabular}}
\caption{Performance comparison for simultaneous shifts with variance increases: Part 2.}
\end{table}

\begin{table}[!htpb]
\resizebox{\textwidth}{!}{
\def\arraystretch{1.5}
\centering
\begin{tabular}{llllllll}
\hline
\hline
 & & & & \multicolumn{4}{c}{EWMA\_GLRT} \\
 \cline{5-8}
\((\mu, \sigma)\) & P\_CUSUM & Wu\_CUSUM & GLRT & $\lambda = 0.01$ & $\lambda = 0.05$ & $\lambda = 0.1$ & $\lambda = 0.2$ \\
\hline
\hline
(0, 1) & 501.7 (5.25) & 490.7 (4.83) & 488.5 (4.7) & 488.5 (4.32) & 505.1 (4.94) & 499.1 (4.74) & 502.8 (4.87) \\  
  (0.25, 0.9) & 91.9 (0.643) & 171.4 (1.48) & 91.2 (0.557) & \textbf{75.9 (0.395)} & 79.6 (0.603) & 106.7 (0.95) & 160.8 (1.52) \\  
  (0.5, 0.9) & 30.9 (0.18) & 37.4 (0.259) & 34.4 (0.187) & 38.7 (0.169) & \textbf{31.5 (0.17)} & 33.7 (0.224) & 46.8 (0.391) \\  
  (0.75, 0.9) & 16.5 (0.083) & 17.2 (0.103) & 17.9 (0.092) & 24.6 (0.104) & 18.4 (0.085) & \textbf{17.7 (0.095)} & 19.7 (0.133) \\  
  (1, 0.9) & 10.6 (0.048) & 10.2 (0.055) & 11.2 (0.055) & 17.1 (0.071) & 12.7 (0.055) & \textbf{11.5 (0.053)} & 11.5 (0.064) \\  
  (1.25, 0.9) & 7.7 (0.032) & 7.1 (0.035) & 7.7 (0.037) & 12.8 (0.052) & 9.3 (0.039) & 8.4 (0.036) & \textbf{7.9 (0.038)} \\  
  (1.5, 0.9) & 6 (0.023) & 5.4 (0.024) & 5.7 (0.026) & 9.9 (0.04) & 7.2 (0.03) & 6.4 (0.027) & \textbf{6 (0.027)} \\  
  (1.75, 0.9) & 4.9 (0.018) & 4.3 (0.017) & 4.5 (0.019) & 7.8 (0.031) & 5.8 (0.023) & 5.1 (0.021) & \textbf{4.8 (0.02)} \\  
  (2, 0.9) & 4.1 (0.015) & 3.6 (0.013) & 3.6 (0.015) & 6.3 (0.026) & 4.7 (0.019) & 4.2 (0.017) & \textbf{3.9 (0.016)} \\  
  (0.25, 0.8) & 64 (0.348) & 214.7 (1.69) & 65.1 (0.291) & 57.7 (0.24) & \textbf{51.8 (0.314)} & 65.2 (0.506) & 109.6 (1.02) \\  
  (0.5, 0.8) & 28.7 (0.145) & 38.8 (0.242) & 32.4 (0.143) & 36.9 (0.143) & \textbf{28.2 (0.129)} & 28.9 (0.165) & 39.4 (0.303) \\  
  (0.75, 0.8) & 16 (0.071) & 17.1 (0.093) & 18.1 (0.081) & 25 (0.095) & 17.8 (0.073) & \textbf{16.8 (0.076)} & 18.5 (0.113) \\  
  (1, 0.8) & 10.5 (0.042) & 10.2 (0.05) & 11.4 (0.05) & 18 (0.069) & 12.7 (0.049) & 11.5 (0.047) & \textbf{11.1 (0.055)} \\  
  (1.25, 0.8) & 7.6 (0.029) & 6.9 (0.031) & 7.8 (0.033) & 13.2 (0.051) & 9.5 (0.037) & 8.5 (0.033) & \textbf{7.9 (0.033)} \\  
  (1.5, 0.8) & 6 (0.021) & 5.2 (0.021) & 5.7 (0.024) & 10.1 (0.039) & 7.3 (0.028) & 6.6 (0.025) & \textbf{6 (0.024)} \\  
  (1.75, 0.8) & 4.9 (0.016) & 4.2 (0.016) & 4.5 (0.018) & 8.1 (0.03) & 5.8 (0.022) & 5.2 (0.019) & \textbf{4.8 (0.019)} \\  
  (2, 0.8) & 4.1 (0.013) & 3.5 (0.012) & 3.6 (0.014) & 6.5 (0.025) & 4.7 (0.018) & 4.3 (0.016) & \textbf{3.9 (0.015)} \\  
  (0.25, 0.7) & 45.3 (0.182) & 246 (1.74) & 49.5 (0.148) & 45.3 (0.167) & \textbf{33.9 (0.157)} & 38.1 (0.233) & 60.9 (0.508) \\  
  (0.5, 0.7) & 25.7 (0.104) & 39.3 (0.212) & 30.6 (0.106) & 34.1 (0.122) & 24.1 (0.093) & \textbf{23.4 (0.112)} & 30.2 (0.209) \\  
  (0.75, 0.7) & 15.3 (0.058) & 17 (0.082) & 18.1 (0.069) & 25 (0.087) & 17.2 (0.062) & \textbf{15.5 (0.06)} & 16.3 (0.088) \\  
  (1, 0.7) & 10.4 (0.037) & 9.9 (0.044) & 11.4 (0.045) & 18.3 (0.066) & 12.8 (0.045) & 11.2 (0.04) & \textbf{10.7 (0.045)} \\  
  (1.25, 0.7) & 7.6 (0.026) & 6.8 (0.027) & 7.8 (0.03) & 13.5 (0.049) & 9.7 (0.034) & 8.5 (0.03) & \textbf{7.8 (0.029)} \\  
  (1.5, 0.7) & 5.9 (0.019) & 5.2 (0.019) & 5.8 (0.021) & 10.4 (0.038) & 7.5 (0.026) & 6.6 (0.023) & \textbf{6 (0.021)} \\  
  (1.75, 0.7) & 4.8 (0.015) & 4.1 (0.014) & 4.4 (0.016) & 8.2 (0.03) & 5.9 (0.021) & 5.3 (0.018) & \textbf{4.8 (0.017)} \\  
  (2, 0.7) & 4 (0.012) & 3.4 (0.011) & 3.6 (0.013) & 6.6 (0.024) & 4.8 (0.017) & 4.3 (0.015) & \textbf{3.9 (0.014)} \\  
     \hline
\end{tabular}}
\caption{Performance comparison for simultaneous shifts with variance decreases: Part 1.}
\end{table}

\begin{table}[!htpb]
\resizebox{\textwidth}{!}{
\def\arraystretch{1.5}
\centering
\begin{tabular}{llllllll}
\hline
\hline
 & & & & \multicolumn{4}{c}{EWMA\_GLRT} \\
 \cline{5-8}
\((\mu, \sigma)\) & P\_CUSUM & Wu\_CUSUM & GLRT & $\lambda = 0.01$ & $\lambda = 0.05$ & $\lambda = 0.1$ & $\lambda = 0.2$ \\
\hline
\hline
(0, 1) & 501.7 (5.25) & 490.7 (4.83) & 488.5 (4.7) & 488.5 (4.32) & 505.1 (4.94) & 499.1 (4.74) & 502.8 (4.87) \\  
   (0.25, 0.6) & 35.1 (0.11) & 282.7 (1.78) & 41.7 (0.102) & 37.1 (0.126) & 25.2 (0.092) & \textbf{24.3 (0.111)} & 32.6 (0.221) \\  
  (0.5, 0.6) & 22.9 (0.076) & 39.4 (0.184) & 29 (0.084) & 31.3 (0.105) & 20.7 (0.07) & \textbf{18.8 (0.073)} & 21.4 (0.122) \\  
  (0.75, 0.6) & 14.6 (0.047) & 16.7 (0.069) & 17.9 (0.058) & 24.3 (0.082) & 16.2 (0.052) & 14.1 (0.047) & \textbf{13.9 (0.059)} \\  
  (1, 0.6) & 10.2 (0.031) & 9.8 (0.038) & 11.5 (0.039) & 18.4 (0.062) & 12.6 (0.041) & 10.9 (0.034) & \textbf{9.9 (0.035)} \\  
  (1.25, 0.6) & 7.5 (0.022) & 6.7 (0.024) & 7.9 (0.026) & 13.9 (0.048) & 9.6 (0.032) & 8.4 (0.027) & \textbf{7.5 (0.025)} \\  
  (1.5, 0.6) & 5.9 (0.017) & 5 (0.016) & 5.8 (0.019) & 10.6 (0.036) & 7.6 (0.025) & 6.7 (0.021) & \textbf{6 (0.019)} \\  
  (1.75, 0.6) & 4.8 (0.013) & 4.1 (0.012) & 4.5 (0.014) & 8.3 (0.029) & 6 (0.02) & 5.3 (0.017) & \textbf{4.8 (0.015)} \\  
  (2, 0.6) & 4 (0.011) & 3.4 (0.01) & 3.6 (0.011) & 6.6 (0.023) & 4.9 (0.016) & 4.4 (0.014) & \textbf{3.9 (0.012)} \\  
  (0.25, 0.5) & 29.2 (0.075) & 328 (1.84) & 37.4 (0.086) & 31.9 (0.104) & 20.3 (0.067) & \textbf{18 (0.063)} & 19.5 (0.098) \\  
  (0.5, 0.5) & 20.8 (0.057) & 39.9 (0.157) & 27.7 (0.071) & 28.9 (0.093) & 18.2 (0.057) & 15.6 (0.049) & \textbf{15.5 (0.064)} \\  
  (0.75, 0.5) & 14 (0.039) & 16.6 (0.06) & 17.8 (0.05) & 23.8 (0.078) & 15.3 (0.047) & 12.9 (0.038) & \textbf{11.8 (0.04)} \\  
  (1, 0.5) & 9.9 (0.027) & 9.7 (0.032) & 11.4 (0.033) & 18.6 (0.061) & 12.4 (0.038) & 10.5 (0.03) & \textbf{9.3 (0.028)} \\  
  (1.25, 0.5) & 7.4 (0.019) & 6.6 (0.02) & 7.9 (0.023) & 14.1 (0.046) & 9.8 (0.029) & 8.4 (0.024) & \textbf{7.4 (0.021)} \\  
  (1.5, 0.5) & 5.9 (0.015) & 4.9 (0.014) & 5.8 (0.017) & 10.7 (0.036) & 7.6 (0.023) & 6.7 (0.019) & \textbf{5.9 (0.017)} \\  
  (1.75, 0.5) & 4.8 (0.012) & 4 (0.011) & 4.5 (0.013) & 8.4 (0.028) & 6.1 (0.018) & 5.3 (0.015) & \textbf{4.8 (0.014)} \\  
  (2, 0.5) & 4 (0.01) & 3.3 (0.008) & 3.5 (0.01) & 6.7 (0.022) & 4.9 (0.015) & 4.3 (0.012) & \textbf{4 (0.011)} \\  
    \hline
\end{tabular}}
\caption{Performance comparison for simultaneous shifts with variance decreases: Part 2.}
\end{table}

We begin our comparison with the \texttt{Wu\_CUSUM} chart, the only other adaptive CUSUM method in our study and thus the most directly comparable to ours. Although designed for detecting changes in multiparameter exponential families, the \texttt{Wu\_CUSUM} chart exhibits notable limitations in the Gaussian case. Most importantly, it cannot detect variance decreases---a key shortcoming in joint monitoring. Even when restricted to detecting mean shifts and variance increases, the \texttt{Wu\_CUSUM} chart generally underperforms relative to ours. In particular, our method consistently achieves shorter \( ARL_1\) for variance increases, typically 1.2 to 1.5 times faster. The sole exception is in detecting large mean shifts, where \texttt{Wu\_CUSUM} holds a slight advantage; however, this advantage is modest (\(ARL_1\) differences are always within 1), whereas our method shows substantially larger gains in the opposite scenarios.

Next, we compare our method with the \texttt{GLRT} chart, often viewed as a performance benchmark due to its joint optimization over the change-point $\tau$ and the unknown parameter space. This comprehensive approach makes the GLRT potentially powerful, but also computationally intensive---especially in sequential or high-dimensional settings. Our simulation results show comparable performance between our method and the GLRT chart in detecting mean shifts only. The GLRT’s clearest advantage lies in detecting variance increases, including in simultaneous shift settings. However, the GLRT’s performance is highly sensitive to tuning parameters---most notably the sliding window size and the lower bound $\gamma$ used in variance estimation (see (\ref{eq: GLRT variance estimate})). This sensitivity is particularly problematic for detecting variance decreases, which the GLRT is not well suited for. Even after optimizing $\gamma$ following the recommendations of \cite{Reynolds2013}, our method consistently outperforms the GLRT in detecting variance decreases, with the largest gains observed when the decrease occurs in isolation. Although the GLRT retains a slight advantage in detecting large mean shifts across all variance scenarios, this benefit comes at the cost of greater computational burden and parameter tuning complexity. These drawbacks limit its practicality in applications where variance decreases are of concern or where ease of implementation is important.

Finally, we compare our proposed chart with the \texttt{EWMA\_GLRT} chart, which---unlike the GLRT chart---does not de-emphasize variance decreases. However, as shown in all tables, the \texttt{EWMA\_GLRT} chart is highly sensitive to the choice of its weighting parameter $\lambda$. While smaller $\lambda$ values are optimal for detecting small shifts and larger values for larger shifts, selecting the appropriate $\lambda$ is difficult without prior knowledge of the shift magnitude. For example, when the OC parameters are $(0, 0.9)$, setting $\lambda = 0.01$ yields an \(ARL_1\) of 162.7---significantly outperforming our method’s \(ARL_1\) of 249.5. However, with $\lambda = 0.2$, the \(ARL_1\) deteriorates to 427.5, representing a substantial drop in performance. This variability highlights a major limitation: in settings of high uncertainty, where the direction or magnitude of shifts is unknown, selecting $\lambda$ becomes highly nontrivial. Consequently, the \texttt{EWMA\_GLRT}'s  performance can degrade rapidly without careful tuning.

Among the control charts considered, no single method universally outperforms all others. While each chart has its strengths, our proposed adaptive CUSUM chart offers a compelling balance of detection power, robustness, and ease of use. Compared to the \texttt{Wu\_CUSUM} chart, it provides broader coverage and superior detection of variance shifts. Against the GLRT chart, it maintains competitive performance with significantly reduced computational and tuning burdens, particularly for variance decreases. Relative to the \texttt{EWMA\_GLRT} chart, it avoids sensitivity to weighting parameter selection while delivering comparable or superior performance. This combination of robustness, adaptability, and efficiency makes our method well suited for practical deployment in uncertain or dynamic monitoring environments.

\section{Concluding Remarks}

In this paper, we develop an adaptive CUSUM chart capable of detecting arbitrary changes in both the mean and variance of a Gaussian process. 
Our proposed chart can be viewed as a natural extension of the adaptive CUSUM chart introduced by \cite{Liu_etal2018}, which was designed to detect mean shifts only. Extending this idea to joint monitoring introduces several new challenges. As in \cite{Liu_etal2018}, our approach involves simultaneously monitoring multiple adaptive CUSUM statistics corresponding to different shift directions. However, instead of two charts, we now require eight---accounting for increases and decreases in both the mean and variance, as well as their combinations. Monitoring all eight statistics simultaneously enables automatic identification of the type of distributional change once an alarm is triggered. However, a novel complication arises in aggregating these eight statistics: unlike the mean-only setting, where the increase and decrease charts are symmetric, the adaptive CUSUM statistics for variance increases and decreases behave quite differently. To address this, we empirically estimate the IC distributions for each adaptive CUSUM statistic, apply transformations to standardize them, and use the maximum of these standardized values as our charting statistic.

One limitation of our method is its reliance on knowledge of the time-varying IC distribution of each adaptive CUSUM statistic. While this is essential for enabling fair comparisons across charting statistics with differing run-length distributions, it can be burdensome in terms of precomputation and storage. Future work could investigate analytical approximations of the run-length distribution---especially in the early monitoring stages---to reduce the need for extensive empirical estimation. Addressing this limitation would further enhance the deployability and real-time practicality of our approach.


\bibliographystyle{plainnat}
\bibliography{bibfile}

\end{document}